\documentclass[prd,showpacs,nofootinbib,preprintnumbers,superscriptaddress]{revtex4}
\usepackage{epsfig}
\usepackage{amssymb}

\newcommand{\cO}{{\cal O}}

\newcommand{\cZ}{{\cal Z}}

\newcommand{\Z}{{Z \!\!\! Z}}
\newcommand{\beqn}{\begin{eqnarray}}
\newcommand{\eeqn}{\end{eqnarray}}
\newcommand{\eq}[1]{(\ref{#1})}

\newcommand{\dd}{\mbox{d}}

\begin{document}

\preprint{KANAZAWA-04-10}
\preprint{ITEP-LAT-2004-14}

\title{Monopole gas in three dimensional SU(2) gluodynamics}

\author{M.~N.~Chernodub}
\affiliation{Institute of Theoretical and  Experimental Physics, B.Cheremushkinskaja 25, Moscow, 117259, Russia}
\author{Katsuya Ishiguro}
\affiliation{Institute for Theoretical Physics, Kanazawa University,
Kanazawa 920-1192, Japan}
\author{Tsuneo Suzuki}
\affiliation{Institute for Theoretical Physics, Kanazawa University,
Kanazawa 920-1192, Japan}

\date{\today}

\begin{abstract}
We study properties of the Abelian monopoles in the Maximal Abelian projection
of the three dimensional pure SU(2) gauge model. We match the lattice monopole
dynamics with the continuum Coulomb gas model using a method of
blocking from continuum. We obtain the Debye screening length
and the monopole density in continuum using numerical results for the
lattice density of the (squared) monopole charges and for the monopole action.
The monopoles treated within our blocking method provide about $75\%$
contribution to the non--Abelian Debye screening length.
We also find that monopoles form a Coulomb plasma which is not dilute.
\end{abstract}
 

\maketitle


\section{Introduction}

According to the dual superconductor mechanism~\cite{DualSuperconductor}
the confinement of quarks in non--Abelian gauge theories is caused by the
monopole--like configurations of the gluonic fields. These configurations
can be identified with the help of the Abelian projection method~\cite{AbelianProjections}. The
basic idea behind this method is to fix partially the
non--Abelian gauge symmetry up to an Abelian subgroup. If the
original non--Abelian gauge group is compact (like the SU(N) group)
then the residual Abelian symmetry group is compact as well.
The compactness of the Abelian group guarantees the existence of the monopoles.

In the low temperature phase of the four dimensional SU(N) gauge model
the monopoles are condensed~\cite{MonopoleCondensation} in agreement with
the dual superconductor mechanism~\cite{DualSuperconductor}.
The condensation leads to the appearance of the dual Meissner effect and,
as a result, to the formation of the chromoelectric string between
fundamental sources of the chromoelectric field (quarks).
Consequently, the quarks get confined. Moreover, the Abelian monopoles are
in the so-called Maximal Abelian projection~\cite{MaA} make
to make a dominant contribution to the zero temperature string tension\cite{AbelianDominance}
(for a review, see Ref.~\cite{Review}).

At the critical temperature, $T=T_c$, the monopole condensate
disappears and at higher temperatures the quarks are no more
confined. The vacuum in the deconfinement phase is filled
by the static monopoles which can not lead to the
confinement of the static quarks. However, the absence of the color
confinement at high temperatures does not mean that the high temperature
physics is perturbative. Indeed, even at $T>T_c$ the quarks
running in the spatial directions are still confined due to
existence of the "spatial string tension" (this is a
coefficient in front of the area term of large spatial Wilson
loops). The spatial string tension is a non--perturbative
quantity which is dominated by contributions of the static
monopoles according\footnote{In a different scenario~\cite{Giovannangeli:2001bh}
the "spatial confinement" problem is suggested to be caused by
magnetic thermal quasi-particles. It seems plausible that these
magnetic excitations are correlated with the Abelian monopoles.}
to Ref.~\cite{AbelianDominanceT}.

The physics of the static monopoles in the high temperature
SU(2) gauge model was investigated in Ref.~\cite{JHEP} using
the method called blocking from continuum (BFC). This method
resembles the idea of the blocking of the continuum {\it
fields} to the lattice~\cite{BlockingOfFields}. In general the
BFC procedure allows us to match lattice results to the continuum
model without going to a deep continuum limit. For example, the BFC method
clearly shows~\cite{JHEP} that in the continuum limit the
static monopoles in the high temperature SU(2) gluodynamics are
described by the three dimensional Coulomb gas model.
Another example is the zero--temperature
4D SU(2) gluodynamics in which the value of the monopole condensate
can be obtained with the help of the BFC method~\cite{CondensateBFC}.
Being encouraged by these results, we apply the BFC
method to the monopoles in the three-dimensional SU(2)
gluodynamics. The confinement mechanism in this model based
on the Abelian monopole dynamics was previously discussed both within
analytical~\cite{ref:three:dim:analyt} and
numerical~\cite{Bornyakov,ref:three:dim:numerical} frameworks.

The plan of the paper is as follows. In Section~\ref{sec:summary} we briefly
recall the results of the Ref.~\cite{JHEP} for
the lattice density of the (squared) monopole charge and for the monopole action expressed via parameters
of the continuum Coulomb gas model. In Section~\ref{sec:numerics} we study the lattice action and the lattice
density of the monopoles in the pure $3D$ SU(2) gauge model. Using
the BFC method we get density of the monopoles and the monopole contribution to the
magnetic screening length in the continuum limit. Our conclusion is presented
in the last Section.

\section{Lattice monopoles from continuum monopoles}
\label{sec:summary}

The standard way to identify the lattice monopoles in Monte
Carlo simulations is to use the DeGrand-Toussaint
construction~\cite{DGT} which calculates the magnetic flux
coming out of lattice 3D cells (cubes). The magnetic charges
obtained in this way are conserved and quantized. The
properties of the lattice monopoles should obviously depend on
the physical size, $b$, of the lattice cell (below we call
these lattice objects as "lattice monopoles of the size $b$").
To get the properties of the monopoles in continuum one should
send the size of the lattice cells to zero, $b \to 0$, what is
usually a difficult numerical problem. The BFC method allows to
get the properties of the monopoles in continuum ("continuum
monopoles") using the results obtained on the lattice with a
finite lattice spacing.

The idea behind the BFC method is to treat each lattice 3D cell
as a "detector" of the magnetic charges of the continuum monopoles.
If the continuum monopole is located inside a
lattice 3D cell then the DeGrand-Toussaint method detects an existence of
a {\it lattice} monopole inside this cell. If the size, $b$, of the
lattice cell is finite then two or more continuum monopoles may be
located inside the cell. The fluctuations of the monopole charges of the
lattice cells must depend on the properties of the continuum monopoles.
As a result, the lattice observables -- such as the vacuum expectation
value of the lattice monopole density (or, the lattice action) -- must carry information
about dynamics of the continuum monopoles. The observables should
depend not only on the size of the lattice cell, $b$, but also on
features of the continuum model which describes the monopole
dynamics.

To avoid misunderstanding we would like to stress from the very
beginning the difference between various lattice monopole sizes $b$.
As we mentioned above, we call the size of the lattice 3D cell -- used
for detection of the monopoles -- as "the size of the lattice
monopole". This should be distinguished from the {\it physical}
radius, $r_0$, of the monopole core~\cite{FiniteRadius} which is obviously
independent on the size of the lattice "detector". In this paper
we disregard the existence of the monopole core and consider the
continuum monopoles as point--like objects. Thus, we get the
finite--sized lattice monopoles by blocking the point--like
continuum monopoles to the lattice.

Below we describe briefly the techniques which appear in the BFC method.
Let us consider a three--dimensional lattice with a finite lattice spacing
$b$ which is embedded in the continuum space-time. The cells of
the lattice are defined as follows:
\beqn
C_s = \Biggl\{b \Bigl(s_i - \frac{1}{2}\Bigr) \leq x_i \leq
b \Bigl(s_i + \frac{1}{2}\Bigr)\,,\quad i=1,2,3 \Biggr\}\,,
\eeqn
where $s_i$ is the lattice dimensionless coordinate and $x_i$
corresponds to the continuum coordinate.

The magnetic charge, $k_s$, inside the lattice cell $C_s$ is
\beqn
k_s = \int\limits_{C_s}\dd^3 x\, \rho(x)\,, \qquad
\rho(x) = \sum_a  q_a \, \delta^{(a)} (x - x^{(a)})\,,
\label{ks}
\eeqn
where $\rho(x)$ is the density of the continuum monopoles, $x_a$
and $q_a$ is the position and the charge (in units of a
fundamental magnetic charge, $g_M$) of $a^{\mathrm{th}}$ continuum
monopole. In three dimensions the monopoles are
instanton--like objects and the monopole trajectories have zero
dimensionality (points). Basic properties of the lattice monopoles
are the same as the ones of the continuum monopoles: the
charge $k_s$ is quantized, $k_s \in \Z$, and conserved in the
three-dimensional sense:
$$
\sum_{s \in \Lambda} k_s \equiv \int\limits_V \dd^3 x \, \rho(x) = 0\,.
$$
Here $\Lambda$ and  $V$ denote the lattice and the continuum volume
occupied by the lattice, respectively.

In the BFC method various properties of the lattice monopoles such as
the lattice monopole density, correlators of the lattice monopole charges,
the effective lattice monopole action {\it etc.} can be calculated in a
continuum monopole model using Eq.~\eq{ks} as a definition of the
lattice monopole charge. In this paper we suppose that the dynamics of
the continuum monopoles is governed by the $3D$ Coulomb gas model:
\beqn
\cZ = \sum\limits_{N=0}^\infty \frac{\zeta^N}{N!}
\Biggl[\prod\limits^N_{a=1} \int \dd^3 x^{(a)} \sum\limits_{q_a = \pm 1}\Biggr]
\exp\Bigl\{ - \frac{g^2_M}{2} \sum\limits_{\stackrel{a,b=1}{a \neq b}}^N
q_a q_b \, D(x^{(a)}-x^{(b)})\Bigr\}\,.
\label{CoulombModel}
\eeqn
The Coulomb interaction in Eq.\eq{CoulombModel} is represented by the inverse
Laplacian $D$, $ - \partial^2_i D(x) = \delta^{(3)}(x)$, and
$\zeta$ is the fugacity parameter.

The model~\eq{CoulombModel} does not exist without properly defined
ultraviolet cut--off. Indeed, the self--energy of the point--like monopoles is a
linearly divergent function. As a result, the fugacity must be renormalized,
$\zeta_{\mathrm{ren}} = \zeta\cdot \exp\{ g^2_m \slash (8 \pi \, r_0)\}$,
where $r_0$ is the ultraviolet cut--off.
In our case this cut--off is given by the size of the monopole core which is of the order of
0.05~fm at zero temperature~\cite{FiniteRadius}. For simplicity we omit the subscript "ren"
in the renormalized fugacity below.

The magnetic charges in the Coulomb gas~\eq{CoulombModel} are screened:
at large distances the two--point charge correlation function is
exponentially suppressed, $\langle \rho(x) \rho(y)\rangle \sim
\exp\{- |x-y| \slash \lambda_D\}$. Here $\lambda_D$ is the Debye
screening length~\cite{Polyakov},
\beqn
\lambda_D = \frac{1}{g_M \sqrt{\rho}}\,,
\label{lambdaD}
\eeqn
which is inversely proportional to the Debye screening
mass, $M_D = \lambda^{-1}_D$.
The density of the continuum monopoles in the leading
order is related to fugacity as~\cite{Polyakov} $\rho = 2 \zeta$.

In Ref.~\cite{JHEP} the lattice monopole action, $S_{mon}(k)$, and
the {\it{v.e.v.}} of the squared magnetic
charge, $\langle k^2_s\rangle$, were calculated starting from the Coulomb
gas model~\eq{CoulombModel}. In the low-density approximation the leading order
contribution to the (squared) density of the monopole charges is~\cite{JHEP}:
\beqn
\langle k^2 (b)\rangle =  \frac{1}{L^3} \, \langle
\sum_{s \in \Lambda} k^2(s) \rangle \equiv
\int_{C_s} \dd^3 x \int_{C_s} \dd^3 y \,
\langle \rho(x) \, \rho(y) \rangle = \rho\, b^3 \, P(M_D \, b)\,,
\label{dens2:latt}
\eeqn
where in the thermodynamic limit (an infinite--volume lattice) the function $P$
is
\beqn
P(\mu) = 1 - \mu^2 \, \int \frac{\dd^3\, q}{{(2\pi)}^3} \,
\frac{1}{q^2 + \mu^2} \, \prod^3_{i=1} {\Biggl[
\frac{2 \sin (q_i \slash 2)}{q_i} \Biggr]}^2
\,.
\label{P}
\eeqn
The finite--volume analogue of Eq.~\eq{P}
can be obtained by the standard substitution:
\beqn
q_i \to \frac{2 \pi k_i}{L_i}\,, \qquad
\int\limits^\infty_{-\infty} \frac{\dd q_i}{2\pi} \to \frac{1}{L_i}
\sum\limits_{k_i \in \Z}\,,
\label{finite:lattice}
\eeqn
where $L_i$ is the lattice size (in units of the lattice spacing)
in $i^{\mathrm{th}}$ direction.

The reason why the density of the squared magnetic charges, $\langle k^2_s\rangle$,
is used instead of the standard definition of the density, $\langle |k_s|\rangle$,
is simple: an analytical treatment of $\langle |k_s|\rangle$ is obviously much
more difficult compared to that of the quantity $\langle k^2_s\rangle$.

The explicit behavior of the lattice monopole density~\eq{dens2:latt} as the function of
$b$ can be obtained in the limit of large lattice
monopoles~\cite{JHEP},
\beqn
\langle k^2 \rangle = C_1 \, \rho \, \lambda_D \, b^2 \cdot \left[1 +
O\left({\left(\lambda_D \slash b \right)}^2\right)\right]\,, \qquad b \gg \lambda_D\,,
\label{TheorDensity:Large}
\eeqn
as well as in the case of small monopoles,
\beqn
\langle k^2 \rangle =
\rho \, b^3 \cdot \left[1 + C_2 \, \rho \, {(b \slash \lambda_D)}^2\,
+ O\left({\left( b \slash \lambda_D\right)}^4\right)\right]\,, \qquad b \ll
\lambda_D\,,
\label{TheorDensity:Small}
\eeqn
where
$C_1 \approx 2.94$ and $C_2 \approx 0.148$.

The proportionality of the density $\langle k^2 \rangle$ to $b^2$
in large--$b$ region has a simple explanation~\cite{JHEP}. In a random
gas of continuum monopoles we would obviously get
$\langle k^2 \rangle \sim \rho \, b^3$. Due to the Debye screening in
the Coulomb monopole gas the monopoles separated from the boundary
of the cell by the distance larger than $\lambda_D$, do not
contribute to $\langle k^2 \rangle$. Consequently, the $b^3$ proportionality
for the random gas turns into $\lambda_D b^2$ in the Coulomb gas and we get
$\langle k^2 \rangle \propto \rho \, \lambda_D \, b^2$.

In the small $b$ region the density of the squared
lattice monopole charges is equal to the density of the continuum
monopoles times the volume of the cell. This is natural, since the
smaller volume of the lattice cell, $b^3$, the smaller chance
for two continuum monopoles to be located at the same cell. Therefore each
cell predominantly contains not more that one continuum monopole, which
leads to the relation $k^2_s = |k_s| = 0,1$. As a result we get
$\langle k^2 \rangle \to \rho_{latt}(b) \to \rho \, b^3$ in the
limit $b \to 0$.

Analogously one can get that in the large-$b$ region the action of the monopole
is given by the lattice Coulomb action~\cite{JHEP},
\beqn
S_{mon}(k) = \frac{1}{\rho\, \lambda_D} \cdot \frac{1}{b^2}
\cdot \sum\limits_{s,s'} k_s\, D_{s,s'}\, k_{s'} + \cdots\,, \qquad
b \gg \lambda_D\,,
\label{TheorAction}
\eeqn
where the coefficient in front of the lattice monopole action is inversely proportional to $b^2$.

\section{Numerical results}
\label{sec:numerics}

We have simulated the pure SU(2) gauge model with the standard Wilson action
$S = - 1/2 \sum_P {\mathrm{Tr}}\, U_P$, where $U_P$ is the plaquette matrix
constructed from the gauge link fields, $U_l$. We have generated
200 configurations of the gauge fields for each chosen value of the
coupling constant,
$\beta = 2.083, 2.5, 3, 3.47, 3.75, 4.5, 5, 6, 6.56, 9, 12, 14.5$, on the
lattice $48^3$. To study the Abelian monopole dynamics we
perform  Abelian projection in the Maximally Abelian (MA) gauge~\cite{MaA}
for each $SU(2)$ configuration. The MA gauge fixing
condition is the maximization of the quantity $R$,
\beqn
\max\limits_{\Omega} R[U^{(\Omega)}]\,,\qquad
R[U] = {\mathrm{Tr}} \sum_{s, \mu} [ U_\mu(s) \sigma_3
U_{\mu}^{\dagger}(s+\hat{\mu}) \sigma_3 ] .
\label{MaA:condition}
\eeqn
under the $SU(2)$ gauge transformations, $U \to U^{(\Omega)} = \Omega^\dagger U
\Omega$. The gauge fixing condition \eq{MaA:condition} is invariant under an Abelian
subgroup of the group of the $SU(2)$
gauge transformations. Thus the condition \eq{MaA:condition}
corresponds to the partial gauge fixing, $SU(2) \to U(1)$.

After the MA gauge fixing, the Abelian, $\{ u_\mu (s)\}$,
and non--Abelian $\{ \tilde{U}_\mu (s)\}$ link fields
are separated:
\beqn
\tilde{U}_\mu (s) = C_\mu (s) u_\mu (s)\,, \quad
 C_\mu (s) = \left(
                 \begin{array}{cc}
                 \sqrt{1-|c_\mu (s)|^2} & -c_{\mu}^{\ast}(s) \\
                 c_{\mu}(s)             & \sqrt{1-|c_\mu (s)|^2}
                 \end{array}
                 \right)\,, \quad
u_\mu (s) =\left(
                 \begin{array}{cc}
                 e^{i \theta_\mu (s)} & 0 \\
                 0                    & e^{-i \theta_\mu (s)}
                 \end{array}
                 \right) .
\eeqn
The vector fields $C_\mu (s)$  and $u_\mu (s)$ transform
like a charged matter and, respectively, a gauge field under the
residual U(1) symmetry. Next we define a lattice monopole current
(DeGrand-Toussaint monopole)~\cite{DGT}. Abelian plaquette
variables $\theta_{\mu\nu}(s)$ are written as \beqn
\theta_{\mu\nu}(s) = \theta_\mu (s) + \theta_\nu (s+\hat{\mu})
                   - \theta_\mu (s+\hat{\nu}) - \theta_\nu (s) \,,
                   \qquad
( -4\pi < \theta_{\mu\nu}(s) \le 4\pi ) .
\eeqn
It is decomposed into two terms using integer variables $n_{\mu\nu}(s)$:
\beqn
\theta_{\mu\nu}(s) \equiv \bar{\theta}_{\mu\nu}(s) + 2\pi n_{\mu\nu}(s)\,,
\qquad ( -\pi < \bar{\theta}_{\mu\nu}(s) \le \pi ) .
\eeqn
Here $\bar{\theta}_{\mu\nu}(s)$ is interpreted as an electromagnetic
flux through the plaquette and $n_{\mu\nu}(s)$ corresponds to the
number of Dirac string piercing the plaquette.
The lattice monopole current is defined as
\beqn
k (s) = \frac{1}{2} \epsilon_{\nu\rho\sigma}
            \partial_\nu n_{\rho\sigma}(s+\hat{\mu})\,.
\label{k}
\eeqn

In order to get the lattice density for the monopoles of various
sizes, $b$, we
perform numerically the blockspin transformations for the lattice monopole
charges. The original model is defined on the fine lattice with
the lattice spacing $a$ and after the blockspin transformation,
the renormalized lattice spacing becomes $b=na$, where $n$ is the
number of steps of the blockspin transformations. The continuum limit
is taken as the limit $a \to 0$ and $n \to \infty$ for a fixed
physical scale $b$.

The monopoles on the renormalized lattices ("extended monopoles",
Ref.~\cite{ExtendedMonopoles}) have the physical size $b^3$.
The charge of the $n$--blocked monopole is equal to the sum of the
charges of the elementary lattice monopoles inside the $n^3$ lattice
cell:
\beqn
k^{(n)} (s) = \sum_{i ,  j ,  l = 0}^{n - 1}
k \bigl(n s + i \hat{\mu} + j \hat{\nu} + l \hat{\rho}\bigr)\,.
\nonumber
\eeqn
For the sake of simplicity we omit below the superscript $(n)$ while referring
to the blocked currents. We perform the lattice blocking with the
factors $n=1 \dots 12$. All dimensional quantities below are measured in
units of the string tension, $\sigma$, the values of which are taken from
Ref.~\cite{Teper:StringTensions,Teper:Glueball}.

\begin{figure}[!htb]
\vskip 3mm
\begin{center}
\includegraphics[scale=0.55,clip=true]{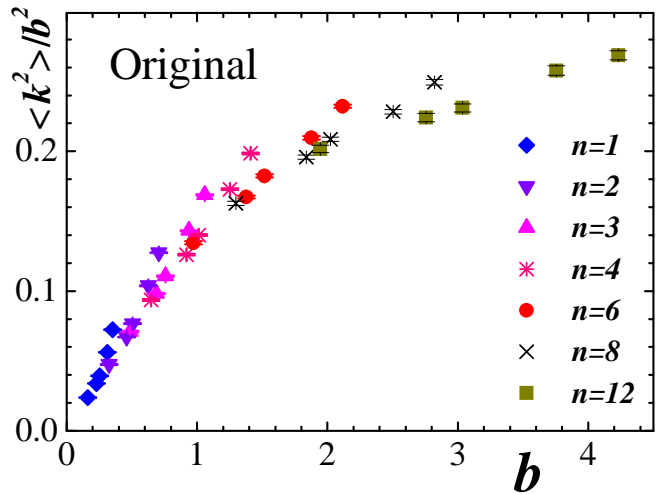}
\end{center}
\vspace{-5mm}
\caption{The density of the squared lattice monopole charges, $\langle k^2_s\rangle$,
divided by $b^2$ as a function of $b$ for various blocking steps $n$
($b$ is given in units of the string tension, $\sigma$).}
\label{fig:Density:Original}
\end{figure}

We show the density of the squared monopole charges (normalized
by the factor $b^2$) in Figure~\ref{fig:Density:Original} as a
function of the scale $b$ for various blocking factors, $n$.
One can easily see from this Figure that the density shows only
an approximate scaling: the data depends not only on $b = a n$,
but also on the fine lattice spacing, $a$, and the blocking
factor, $n$, separately. The violation of the $b$--scaling --
being much bigger then the error bars of the data -- strongly
affects the physical results.

The origin of the $b$--scaling violation is in the presence of
the ultraviolet lattice artifacts which express themselves in
the form of the tightly bound monopole--anti-monopole pairs
(magnetic dipoles). These dipoles are living on the fine
lattice and typical distance between the constituents of the
these dipoles is of the order of the fine lattice spacing, $a$.
Thus, in the continuum limit these dipoles must disappear.
However, on the finite lattices they affect the results
drastically.

In order to get rid of the ultraviolet artifacts we have
removed the tightly--bound dipole pairs from all configurations
using a simple numerical algorithm. Namely, we remove a
magnetic dipole if it is made of a monopole and an
anti-monopole which are touching each other ({\it i.e.}, this
means that the centers of the corresponding cubes are located
at the distance smaller or equal than $\sqrt{3} a$). Note, that
we first apply this procedure to the elementary
$a^3$--monopoles, and only then we perform the blockspin
transformations. Below we discuss the results obtained for the
monopole ensembles with the artificial UV--dipoles removed.

\begin{figure}[!htb]
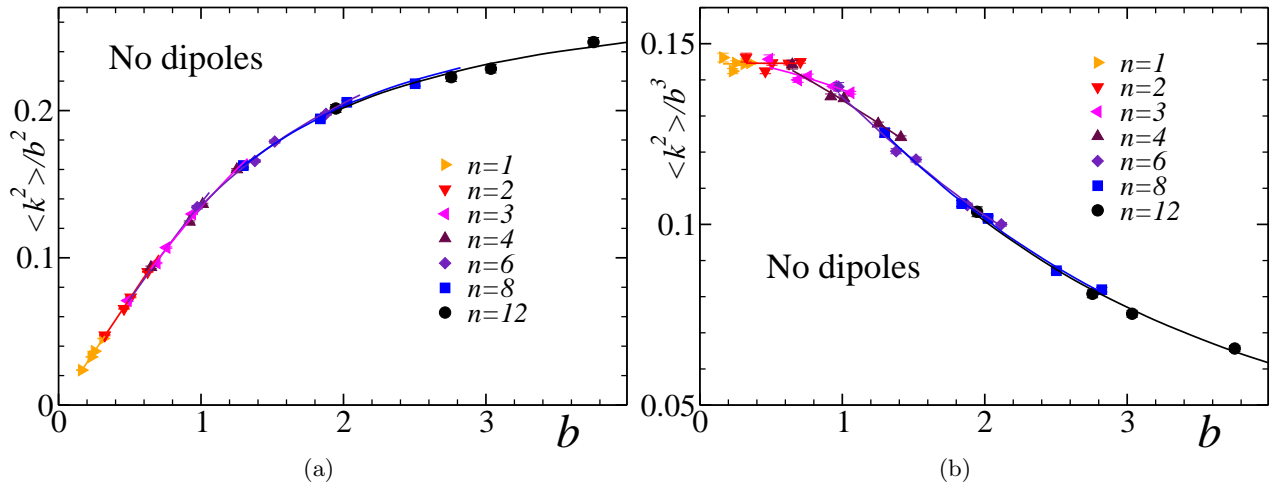

\begin{center}
\begin{tabular}{cc}
\includegraphics[scale=0.35,clip=true]{rho2b.eps}  &
\includegraphics[scale=0.35,clip=true]{rho3b.eps} \\
(a) & (b) \\
\end{tabular}
\end{center}
\caption{The density of the squared monopole charges, $\langle k^2_s\rangle$,
with the UV dipoles removed. The density is normalized (a)
by $b^2$ and (b) by $b^3$. The fits by the function~\eq{dens2:latt} are shown by
dashed lines for each value of the blocking step, $n$.}
\label{fig:Density:NoDipoles}
\end{figure}
In Figure~\ref{fig:Density:NoDipoles}(a) we show the density of the
squared monopole charges as a function of the scale $b$ for various values of the
blocking factor, $n$. The density is normalized by the factor $b^2$.
One can see that the $b$-scaling violations are very small. As the blocking size $b$
increases the slope of the ratio $\langle k^2_s\rangle/b^2$ becomes less steep.
This behavior is in a qualitative agreement with the prediction from the Coulomb
gas model~\eq{TheorDensity:Large} which says that in the high--$b$ limit the
ratio $\langle k^2_s\rangle/b^2$ should converge to a constant.

On the other hand the quantity $\langle k^2_s\rangle$ should be proportional to
$b^3$ in the small--$b$ region~\eq{TheorDensity:Small}.
One can indeed observe from Figure~\ref{fig:Density:NoDipoles}(b)
that the ratio $\langle k^2_s\rangle/b^3$ does tend to a constant
at small $b$. Note that there is a small scaling
violation in this region which is due to the presence of the lattice artifacts at the
scale $b \sim a$. In order to get artifact--free results we will use below
large--$b$ monopoles.

The values of the parameters of the Coulomb gas model in the
continuum limit, Eq.~\eq{CoulombModel}, can be obtained by fitting
the numerical results for $\langle k^2_s\rangle$ by the theoretical
prediction~(\ref{dens2:latt}),(\ref{finite:lattice}). Technically, for each
value of the blocking step, $n$, we have a set of the data corresponding to
different values of the lattice coupling $\beta$, and, consequently, to different
values of $b = n\cdot a(\beta)$. Note that by fixing $n$ we simultaneously fix
the extension of the coarse lattice, $L/n$, in units of $b$. The size of the
coarse lattice enters in Eq.~\eq{finite:lattice}. We fit the set of the data for the
fixed blocking step $n$. The best fit curves are shown in
Figures~\ref{fig:Density:NoDipoles}(a) and (b) by dashed lines. The quality of the
fit is very good, $\chi^2/d.o.f. \sim 1$.

\begin{figure}[!htb]
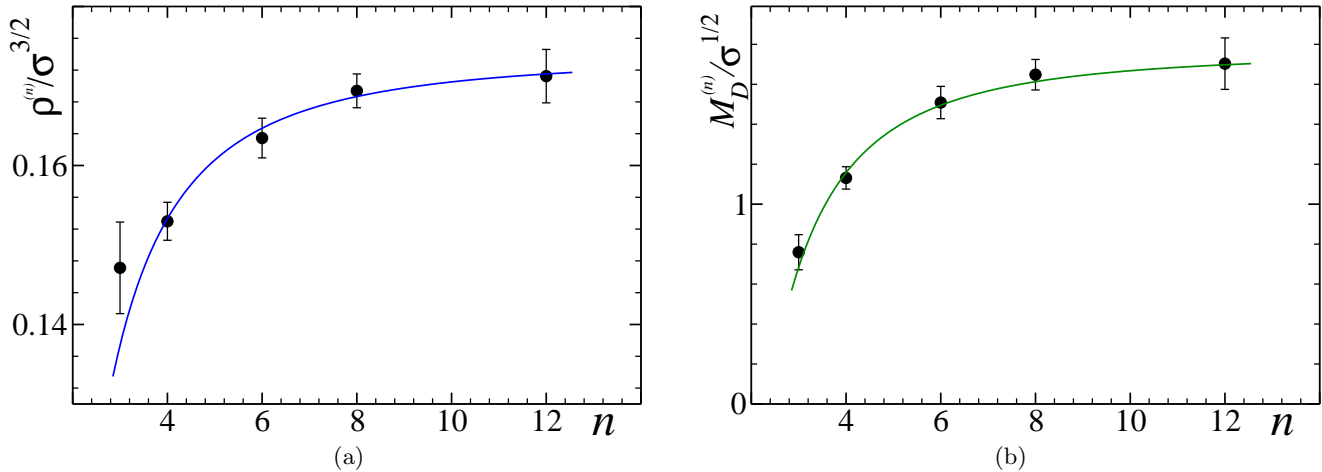

\begin{center}
\begin{tabular}{cc}
\includegraphics[scale=0.35,clip=true]{rho.eps} \hspace{5mm} &
\includegraphics[scale=0.35,clip=true]{m.eps} \\
(a) & (b) \\
\end{tabular}
\end{center}
\caption{(a) The density of the continuum monopoles, $\rho$,
and (b) the Debye screening mass, $M_D$,
obtained with the help of the fits of the $n$--blocked squared
monopole density by function~\eq{dens2:latt}. The large--$n$
extrapolation~\eq{eq:large:n} is shown by solid lines.}
\label{fig:RhoM}
\end{figure}

The density fits give the values of the continuum monopole
density, $\rho^{(n)}$, and the Debye mass, $M^{(n)}_D$, which are shown
in Figures~\ref{fig:RhoM}(a,b) as a function of blocking step $n$. All
these results are given in units of the string tension.
The influence of the finite lattice spacing should vanish in the limit of large $b$, or,
in our case, in the limit of large $n$: $\lim_{n\to\infty} \cO^{(n)} = \cO^{\mathrm{ph}}$,
where $\cO$ stands for either $\rho$ or $M^D$ and the superscript "ph" indicate the artifact--free
physical value. We found that for $n >2$  the dependence of both $\rho$ and $M_D$ on the
blocking size $n$ can be approximately described as
\beqn
\cO^{(n)} = \cO^{\mathrm{ph}} + {\mathrm{const}} \cdot n^{-2}\,,
\label{eq:large:n}
\eeqn
The extrapolation~\eq{eq:large:n} is shown in Figure~\ref{fig:RhoM}.
We get the physical values for the monopole
density $\rho$ and the Debye screening mass $M_D$ coming from the Coulomb
gas model (here and below we omit the superscript "ph" for the extrapolated values):
\beqn
\rho \, / \sigma^{3/2}= 0.174(2)\,, \qquad
M_D \, / \sigma^{1/2}= 1.77(4)\,.
\label{eq:Num:RhoM}
\eeqn
The value of $M_D$ may be treated as the "monopole contribution to
the Debye screening mass".

\begin{figure}[!htb]
\begin{center}
\includegraphics[scale=0.4,clip=true]{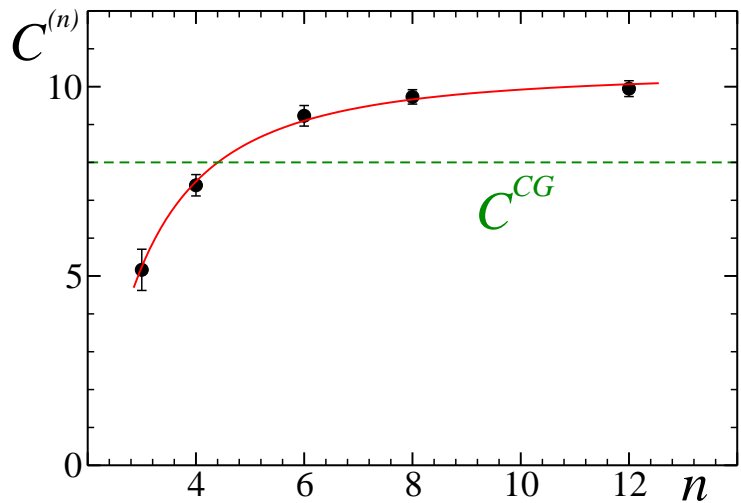}
\end{center}
\caption{The same as in Figure~\ref{fig:RhoM} but for the ratio~\eq{eq:Csp}.
The dashed line corresponds to the low density limit of Coulomb Gas
model~\cite{Polyakov}, $C^{\mathrm{CG}}=8$.}
\label{fig:Csp}
\end{figure}

The self--consistency check of our approach can be done with the help of the quantity
\beqn
C = \frac{M_D\, \sigma}{\rho}\,,
\label{eq:Csp}
\eeqn
which is known to be equal to eight ($C^{\mathrm{CG}}=8$)
in the low density limit of the Coulomb gas model~\cite{Polyakov}.
In Figure~\ref{fig:Csp} we plot our numerical result for $C$
as a function of $n$.

The large--$n$ extrapolation~\eq{eq:large:n} gives
\beqn
C = 10.1(1)\,, \qquad {\it i.e.} \qquad
\frac{C}{C^{\mathrm{CG}}} = 1.26(3)\,.
\label{eq:Num:Csp}
\eeqn
The quantity $C$ is about $25\%$ larger than the one predicted by the Coulomb gas
model in the low monopole density approximation, $C^{\mathrm{CG}}_{sp}=8$.
The discrepancy is most likely explained by the invalidity of the assumption
that the monopole density is low. Indeed, the low--density approach requires for
the monopole density to be much lower than a natural scale for the
density, $g^6$ (remember that the coupling $g$ has the dimensionality $mass^{1/2}$).
The requirement $\rho \ll g^6$ can equivalently be reformulated as
$\rho/M_D^3 \gg 1$, which means that the number of the monopoles in a unit Debye volume,
$\lambda_D^3 \equiv M_D^{-3}$, must be high. Taking the numerical values
for $\rho$ and $M_D$ from Eq.~\eq{eq:Num:RhoM} we get: $\rho/M_D^3 \approx 0.03 \ll 1$.
Thus, the low--density assumption is not valid in the 3D SU(2) gluodynamics.
However, the discrepancy of $25\%$ observed in the quantity $C$,
Eq.~\eq{eq:Num:Csp}, is a good signal that the Coulomb gas model may still provide
us with the predictions valid up to the specified accuracy.

One can compare our result for the monopole density, Eq.~\eq{eq:Num:RhoM}, with the
result obtained by Bornyakov and Grigorev in Ref.~\cite{Bornyakov}, $\rho^{\mathrm{BG}} = 2^{-7} (1 \pm 0.02) \, g^6$. Using the
result of Ref.~\cite{Teper:StringTensions}, $\sqrt{\sigma} = 0.3353(18) \, g^2$, we get
the value $\rho^{\mathrm{BG}} / \sigma^{3/2} = 0.207(5)$, which is close to our
{\it independent} estimation in the continuum
limit~\eq{eq:Num:RhoM}: $\rho/\rho^{\mathrm{BG}} = 0.83(4)$. The result of
Ref.~\cite{Bornyakov} is about $20\%$ higher than our estimation for the monopole density.
Thus, although the condition of the low monopole density approximation is strongly violated,
the BFC method (based on the dilute gas approximation) gives the value of the monopole density
which is consistent with other measurements.

It is interesting to compare the result for the screening mass~\eq{eq:Num:RhoM}
with the lightest glueball mass measured in
Refs.~\cite{Teper:StringTensions,Teper:Glueball},
$M_{O^{++}} = 4.72(4)\, \sqrt{\sigma}$. In the Abelian picture, the
mass of the ground state glueball obtained with the help of the correlator,
$$\langle F^2_{\mu\nu}(0)\, F^2_{\alpha\beta}(R)\rangle =
{\mathrm{const.}}\, e^{- M_{O^{++}} \, R} + \dots\,,$$
must be twice bigger than the Debye screening mass, $2 M_D / M_{O^{++}} = 1$,
where the Debye mass is given by the following correlator
$$\langle F_{\mu\nu}(0)\, F_{\mu\nu}(R)\rangle = {\mathrm{const.}}\, e^{- M_{D} \, R} + \dots\,.$$
The comparison of our result~\eq{eq:Num:RhoM} with the result of
Refs.~\cite{Teper:StringTensions,Teper:Glueball} gives $2 M_D/M_{O^{++}} = 0.75(4)$. The
deviation is of the order of $25\%$ similarly to case of the quantity $C$.

Let us also compare our result for the monopole contribution to the Debye screening
mass, Eq.~\eq{eq:Num:RhoM}, with the direct measurement of the Debye
mass in 3D SU(2) gauge model made in
Ref.~\cite{Karsch}, $m^{SU(2)}_D/\sqrt{\sigma} = 1.39(9)$. The values agree with
each other within the $25$ per cent: $m_D/m^{SU(2)}_D = 1.27(11)$.
Approximately the same accuracy is observed in the four--dimensional SU(2) gauge
theory for the monopole contribution to the fundamental string tension~\cite{ref:Bali}.

Finally, we have made the cross--check of our result by a numerical calculation of the effective
monopole action. We used an inverse Monte-Carlo algorithm described in Ref.~\cite{Yazawa}
and we used the following form of the trial action:
\beqn
S_{mon}(k) = f_1 \cdot \sum\limits_{s} k^2_s + f_{Coul} \cdot \sum\limits_{s,s'} k_s\, D_{s,s'}\, k_{s'}\,.
\label{NumericalAction}
\eeqn
The choice of this type of the trial action is motivated by the following reasons. {}From the point of view
of reliability of the results the data is most valuable in the large-$b$ region since the subtraction of the
ultraviolet monopoles can not change the infrared physics. In this region
the action is expected to be given by the Coulomb term~\eq{TheorAction}. On the other hand the subtraction of the
ultraviolet monopoles must affect the ultraviolet (or, local) terms in the monopole action. Thus we added
to the trial action~\eq{NumericalAction} the most local term with the coupling $f_1$. The role of this term is to
take into account the effects which are caused by the monopole subtraction.

\begin{figure}[!htb]
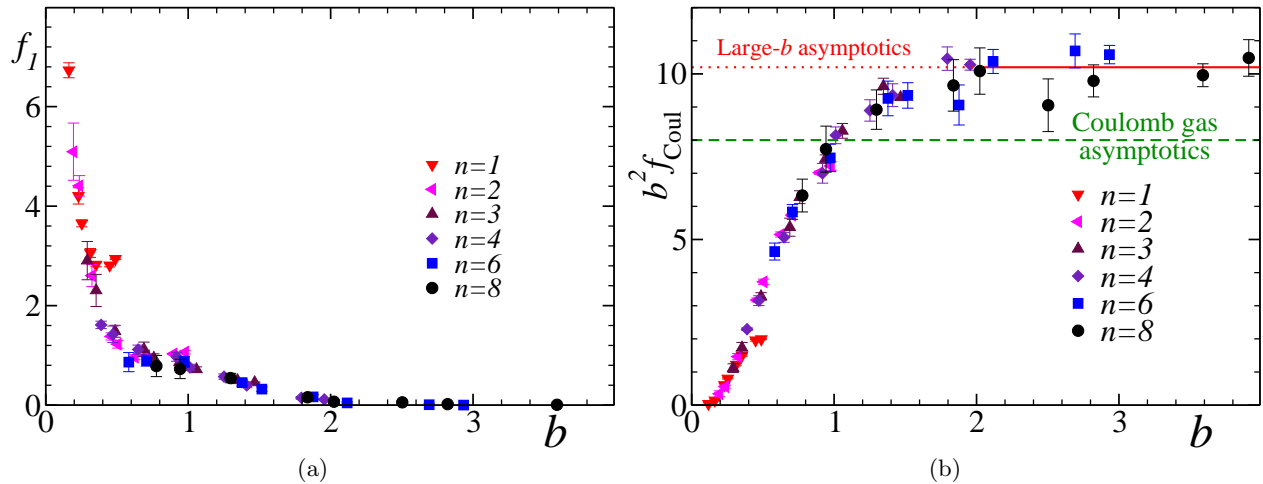

\begin{center}
\begin{tabular}{cc}
\includegraphics[scale=0.35,clip=true]{f1.eps}  &
\includegraphics[scale=0.35,clip=true]{fcoul.eps} \\
(a) & (b) \\
\end{tabular}
\end{center}
\caption{The coupling $f_1$ and the product $b^2\,f_{Coul}$ of the action~\eq{NumericalAction}
$vs.$ $b$ for each value of the blocking step, $n$.}
\label{fig:Couplings}
\end{figure}

We depict the coupling $f_1$ and the product $b^2\,f_{Coul}$ as a function of $b$ in
Figures~\ref{fig:Couplings}(a,b), respectively.
The strong coupling region with $\beta \leqslant 3.0$ was not included into these figures due to
large finite size effects. Indeed, in this case $a \sqrt{\sigma} \gtrsim 0.6$ which implies
that $a M_D \approx 1$ (here we used the value of $M_D \equiv \lambda_D^{-1}$ obtained from
the fits of the monopole density~\eq{eq:Num:RhoM}).

Note that small--$n$ results are more sensitive to the subtraction of the ultraviolet magnetic dipoles.
This is the reason for noticeable $b$--scaling violations around $b\sim 0.5 \sigma^{-1/2}$
for $n = 1$ in both Figures~\ref{fig:Couplings}(a,b).

According to the analytical prediction~\eq{TheorAction} the product of the coupling $f_{Coul}$ and $b^2$
should tend to a constant in the limit $b \gg \lambda_D$
while the $f_1$--term should disappear from the action. Numerically, the coupling $f_1$ is diminishing  as $b$ increases,
as seen in Figure~\ref{fig:Couplings}(a). One can also check numerically that this coupling is vanishing quicker than $1/b^2$ at
$b \gtrsim 1.5 \sigma^{-1/2}$. So in the large--$b$ region we are indeed left only with the lattice Coulomb term in the
action. If the scale $b$ is expressed in units of the string tension then the product $b^2 f_{Coul}(b)$ tend to the
quantity~\eq{eq:Csp}. Figure~\ref{fig:Couplings}(b) shows
that large-$b$ plateau in distribution of  $b^2 f_{Coul}(b)$ starts from $b \gtrsim 2 \sigma^{-1/2}$. Averaging the available
data with $b \geqslant 2 \sigma^{-1/2}$ we get $C = 10.2(3)$ which is in an excellent agreement with the result~\eq{eq:Num:Csp}
obtained from the monopole action.

\section{Conclusions}

We have shown that the dynamics of the Abelian monopoles in the
three--dimensional SU(2) gauge model can be described by the
Coulomb gas model. Using a novel method, called the blocking of
the monopoles from continuum, we calculated the monopole
density and the Debye screening mass in continuum,
Eq.~\eq{eq:Num:RhoM}, using the numerical results for the
(squared) monopole charge density. The self-consistency of the results
was checked by the independent analysis of the lattice monopole action.
We conclude that the Abelian monopole gas in the 3D SU(2) gluodynamics is not
dilute. Nevertheless, the continuum values of the monopole density ($\rho
= 0.174(2)\, \sigma^{3/2}$) and the Debye screening mass ($M_D =
1.77(4)\, \sigma^{1/2}$) -- obtained with the help of the dilute
monopole gas model -- are consistent within the accuracy of $25\%$
with the known data obtained from independent measurements.

\begin{acknowledgments}
This work is supported by JSPS Grant-in-Aid for Scientific
Research on Priority Areas 13135210,  (B) 15340073, JSPS grant S04045, and
grants RFBR 01-02-17456, DFG 436 RUS 113/73910, RFBR-DFG
03-02-04016 and MK-4019.2004.2. The numerical simulations have been
performed on NEC SX-5 at RCNP, Osaka University.
\end{acknowledgments}

\end{document}